\documentclass[a4paper,fleqn,usenatbib]{mnras}

\usepackage{mathptmx}
\usepackage[T1]{fontenc}
\usepackage{ae,aecompl}

\usepackage{graphicx}	
\usepackage{amsmath}	
\usepackage{amssymb}	
\usepackage[dvipsnames]{xcolor}
\usepackage[normalem]{ulem}
\title[Neutron star crust cooling in KS 1731-260]{Neutron star crust cooling in KS 1731-260: the influence of accretion outburst variability on the crustal temperature evolution}
\author[L. S. Ootes et al.]{
Laura S. Ootes,$^{1}$\thanks{\vspace{-0.4cm}E-mail: l.s.ootes@uva.nl (LSO)}
Dany Page,$^{2}$
Rudy Wijnands,$^{1}$
and Nathalie Degenaar$^{1, 3}$
\\
$^{1}$Anton Pannekoek Institute for Astronomy, University of Amsterdam, Postbus 94249, 1090 GE Amsterdam, The Netherlands\\
$^{2}$Instituto de Astronom\'{i}a, Universidad Nacional Aut\'{o}noma de M\'{e}xico, M\'{e}xico, D.F. 04510, M\'{e}xico\\
$^3$Institute of Astronomy, University of Cambridge, Madingley Road, Cambridge CB3 OHA, United Kingdom
}

\date{Accepted 2016 June 28. Received 2016 June 6; in original form ZZZ}

\pubyear{2016}

\begin{document}
\label{firstpage}
\pagerange{\pageref{firstpage}--\pageref{lastpage}}
\maketitle

\begin{abstract}
Using a theoretical model, we track the thermal evolution of a cooling neutron star crust after an accretion induced heating period with the goal of constraining the crustal parameters. We present for the first time a crust cooling model -- \textit{NSCool} -- that takes into account detailed variability during the full outburst based on the observed light curve. We apply our model to KS 1731-260. The source was in outburst for $\sim$12 years during which it was observed to undergo variations on both long (years) and short (days-weeks) timescales. Our results show that KS 1731-260 does not reach a steady state profile during the outburst due to fluctuations in the derived accretion rate. Additionally, long time-scale outburst variability mildly affects the complete crust cooling phase, while variations in the final months of the outburst strongly influence the first $\sim$40 days of the calculated cooling curve. We discuss the consequences for estimates of the neutron star crust parameters, and argue that detailed modelling of the final phase of the outburst is key to constraining the origin of the shallow heat source.
\end{abstract} 




\section{Introduction}
Neutron stars in low mass X-ray binaries (LMXBs) accrete matter from their companions. Some sources accrete persistently, while others accrete episodically (transient sources). The length of an accretion episode is typically weeks to months. However, in quasi-persistent sources the accretion outburst lasts for $\gtrsim$1 year. 

Large amounts of gravitational potential energy are released as matter falls onto the neutron star surface. The observed surface temperatures during accretion are of the order $10^7$ K. If the interior temperature is lower, heat generated at the surface may flow inward. The accreted material is fused into heavier elements in the underlying ocean, producing heat that can raise the ocean's temperature well above the surface temperature. This stops the inflow of gravitationally generated heat \citep{Fujimoto:1984aa,Miralda-Escude:1990aa}, while a fraction of the thermonuclear heat may still flow inward as long as the temperature gradient from the ocean towards deeper layers is negative. Finally, the accreted material is compressed to higher densities where electron captures, neutron emission, and pycnonuclear reactions result in the release of $\sim$$1-2$ MeV per accreted nucleon \citep{Sato:1979pd,haensel1990}. Most of this energy is released deep in the crust at densities $10^{12-13}\text{g cm}^{-3}$ where the pycnonuclear reactions take place. This energy heats up the region where the reactions occur \citep{Miralda-Escude:1990aa} and slowly flows into the stellar core. 

This last mechanism is known as the deep crustal heating model  and has been invoked to explain the observed high quiescent luminosities of neutron stars in transient LMXBs \citep{brown1998}. Even with short, but repeating, accretion outbursts deep crustal heating will heat up the core of the star until it reaches an equilibrium temperature 
\citep{brown1998,colpi2001}. 

In quasi-persistent sources (and some normal transients), the effect of the deep crustal heating is strong enough that the crust is brought out of thermal equilibrium with the core \citep{rutledge2002}. For an outburst with constant accretion rate that lasts years to decades, a steady state will be reached in the crust in which the amount of heat released in the crust is high enough to compensate the heat loss to the core and the surface \citep{brown2009, page2013}. Once the system returns to quiescence, the crust cools down until thermal equilibrium with the core is restored.

The time-scale of crustal heating and cooling depends on crustal microphysics (e.g., thermal conductivity and specific heat, which depend strongly on chemical composition), and macroscopic parameters (crust mass and thickness), as well as the outburst properties: accretion rate and outburst duration. During quiescence, the core temperature sets the quiescent base level. \citet{rutledge2002} proposed that observations during the crust cooling period allow one to probe the crustal physics of neutron stars. To date, crust cooling after an accretion outburst has been observed in eight sources \citep[e.g.][]{wijnands2001, wijnands2003, degenaar2013, degenaar2014, degenaar2015, fridriksson2010, homan2014, waterhouse2016}. 

During the past 15 years various crust cooling models have been developed that explain the observed cooling curves based on the heat diffusion equations, which allows one to quantify the properties of the neutron star crust \citep[e.g.][]{shternin2007, brown2009, page2013, horowitz2015, turlione2015}. While it was initially thought that accreting neutron stars would have amorphous crusts \citep{schatz1999, brown2000}, modelling the observed cooling curves has revealed that they have relatively low impurity crusts instead, because the fast overall crust cooling time-scale can only be explained by high crustal conductivity \citep[e.g.][]{shternin2007, brown2009}. Modern simulations of the crustal structure reconciled these high conductivities with theory and moreover, found that the impurity parameter is not constant throughout the crust, demonstrating a significant difference between the outer and inner crust \citep{horowitz2007, horowitz2015, mckinven2016}. 

Comparing cooling observations with calculated cooling curves has also encountered a new problem. While for some sources the cooling can be modelled using the heat sources expected within the deep crustal heating model \citep[e.g.][]{page2013,degenaar2015}, others are observed to be hotter than predicted in the earliest phase of the crust cooling period. The cooling observations of these sources can only be explained if there is an additional heat source in the outermost layers of the neutron star crust, which has been referred to as the shallow heat source \citep[e.g.][]{brown2009, degenaar2011b, page2013, turlione2015,deibel2015, waterhouse2016}. For those sources that require additional shallow heat, the specific amount differs per source: in most sources $1-2 \text{ MeV nuc}^{-1}$ is enough to explain the observations, while one source requires $\sim$$10$ MeV nuc$^{-1}$ \citep{deibel2015}. It should be noted that additional heat sources at shallow depths in the crust were also proposed to explain certain observed phenomena during thermonuclear burning on accretion neutron stars \citep[e.g.][]{linares2012, zand2012, deibel2016}. Searches to find the origin of the shallow heat source are still in preliminary stages \citep[e.g.][]{medin2014}, and many of the suggested mechanisms are unable to provide additional heat up to 10 MeV nuc$^{-1}$ \citep[see discussion by][]{deibel2015}.

Previous crust cooling studies modelled the outburst profile as a step function, assuming a constant accretion rate during the outburst. However, observed outburst light curves generally show large variations in X-ray luminosity, indicating that the accretion rate is not constant. Moreover, such step functions do not take into account any decrease in accretion rate during the last phase of the outburst when the source returns to quiescence. So far only two models have been presented in which accretion rate decay was taken into account \citep{page2013,deibel2015}. 

In this letter we present for the first time a model that takes into account outburst variability on both long and short time-scales, using the observed outburst profile. We apply our model to KS 1731-260. The source was in outburst for over 12 years when it returned to quiescence in early 2001 \citep{wijnands2001,wijnands2002}. We chose this source because of its long observed cooling curve \citep{cackett2006,cackett2010}, and because observations were carried out throughout the whole outburst. However, the number of observations in the first $\sim$7 years, as well as in the last two months of the outburst is limited. We determine the effect of accretion fluctuations on the evolution of the thermal state of the neutron star crust during the outburst and discuss the influence on the calculated cooling curves and the consequences for constraining neutron star crustal parameters. Additionally, we consider the uncertainties yielded by the limitations of the available outburst data.

\vspace{-0.1cm}\section{Crust cooling model}
We model the thermal evolution of the neutron star crust during and after the outburst using an improved version of the cooling code \textit{NSCool} \citep{page2013}. The code solves the thermal evolution equations (energy transport and conservation) taking into account general relativistic effects. We make use of the A18+$\delta$v+UIX* equation of state \citep{akmal1998} for the core and assume the original catalysed crust to be fully replaced by accreted material as described by \citet{haensel2008}. This model assumes the initial composition of the nuclear-burning ashes to be $^{56}$Fe. The code integrates the thermal profile of the neutron star up to the outer boundary, defined at a density $\rho_b=10^{8} \text{ g cm}^{-3}$. To calculate the effective temperature $T_\text{eff}$ from the boundary temperature $T_\text{b}(\rho_\text{b})$ we use the accreted envelope models from \citet{potekhin1997}, where we allow the amount of light elements in the envelope (parametrised through their column density $y_\mathrm{light}$) to be adjustable.

We assume that $1.93$ MeV nuc$^{-1}$ is released in the crust during an accretion outburst \citep{haensel2008}. An additional shallow heat source is defined with strength $Q_\text{sh}$ between $\rho_\text{sh,min}$ and $\rho_\text{sh,max}$, where $\rho_\text{sh,max}=5\rho_\text{sh,min}$ \citep[similar to][]{deibel2015}. The released amount of shallow heat is proportional to the accretion rate. We neglect the contribution of superfluid phonons to the crustal conductivity and only consider electron contribution. Electron scattering by ions and phonons is calculated following \citet{gnedin2001} and by impurities as in \citet{yakovlev1980}. Impurities in the crust are parameterised by the impurity parameter $Q_{imp}$. Finally, we set a uniform (red-shifted) temperature $T_0$ in the stellar interior prior to the onset of accretion as initial condition.

\vspace{-0.2cm}
\subsection{Modelling the accretion outburst}
\begin{figure}	
     \includegraphics[width=\columnwidth]{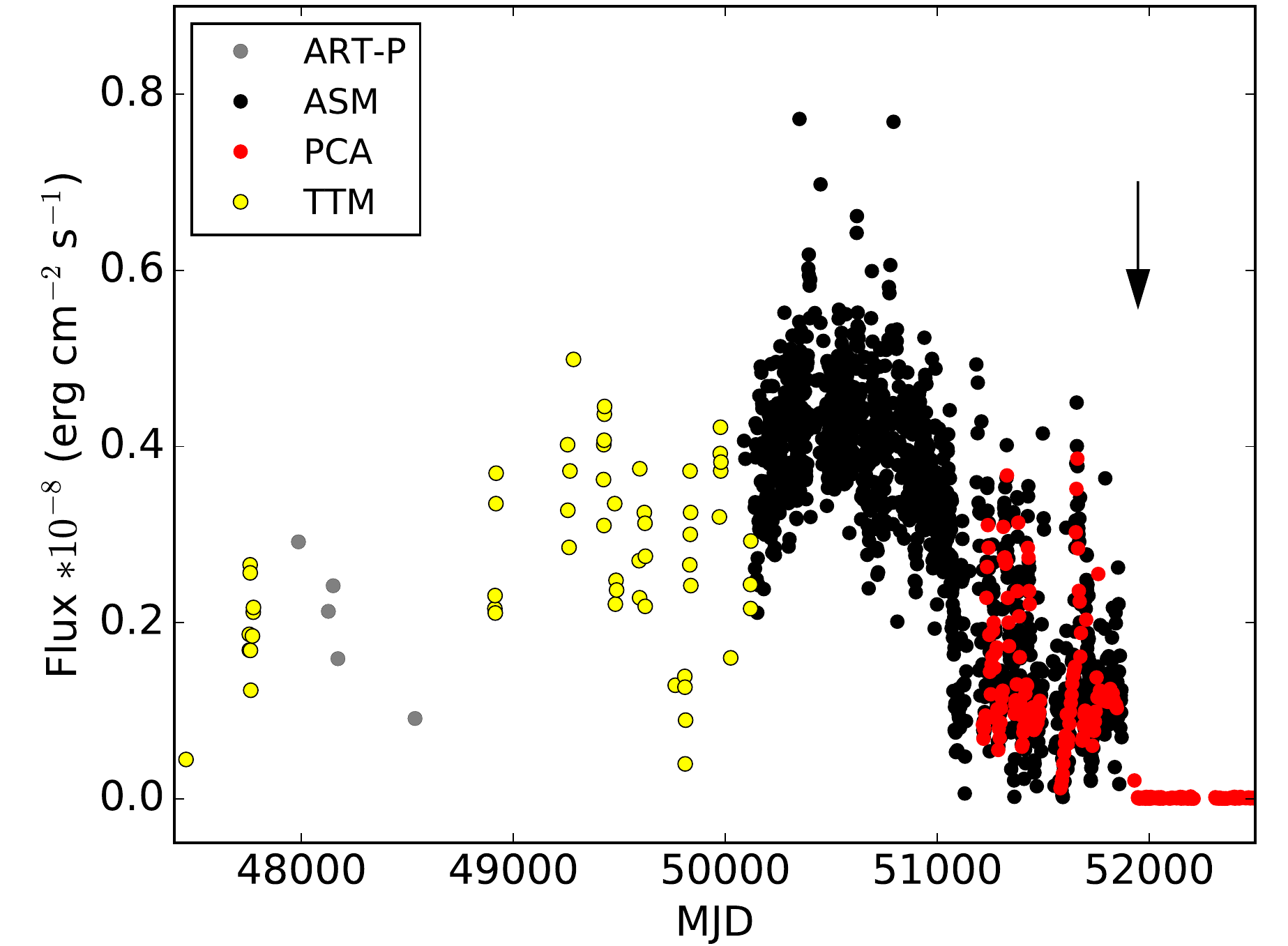}
    \vspace{-0.2cm} \caption{Light curve of KS 1731-260 during its 12 year outburst in 2-10 keV flux units. The arrow indicates the day for which the first non-detection of the source was obtained. All points after this day are non-detections.}
     \label{fig:lightcurve}
\end{figure}
\begin{figure*}
     \includegraphics[width=1.05\columnwidth]{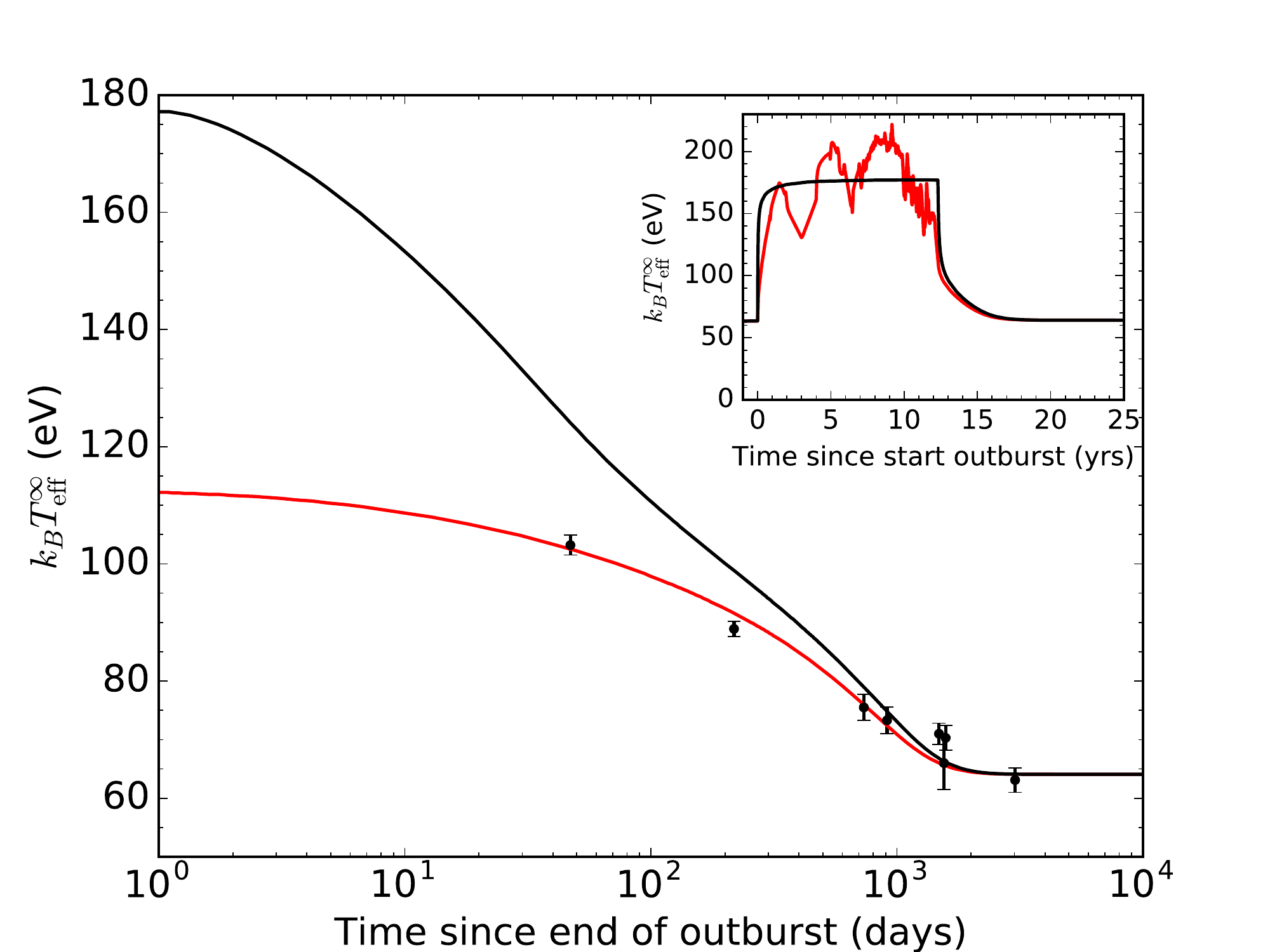}\qquad\includegraphics[height=0.8\columnwidth]{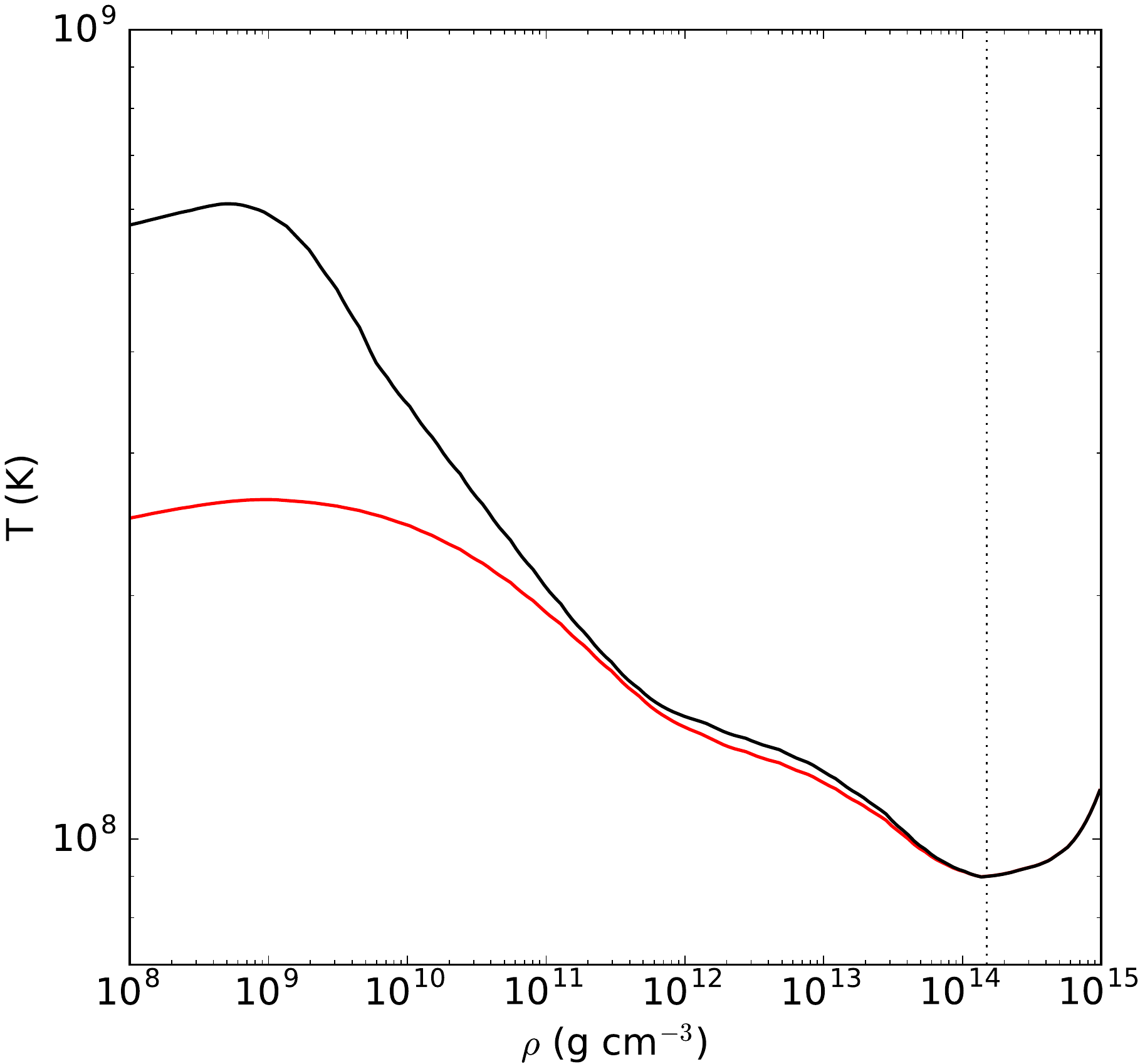}
     \caption{Left: Calculated cooling curves for KS 1731-260 from model 1 (red, assuming a time variable accretion rate estimated from the observed outburst behaviour), and model 2 (black, assuming constant accretion rate). The inset shows the effective temperature during and after the accretion outburst as function of time since the start of the outburst. Right: Calculated temperature profiles in the crust  up to the boundary density $\rho_b=10^8 \text{ g cm}^{-3}$ for models 1 and 2 at the end of the outburst. The dotted line indicates the crust-core boundary. Note that the temperature in the right panel is the local, i.e., non-redshifted, temperature.}\vspace{-0.4cm}
     \label{fig:step_real}
\end{figure*}
To take into account accretion variability during the outburst, we determine a time-dependent accretion rate based on the observational data. KS 1731-260 was discovered in August 1989 with \textit{Mir-Kvant}, but reanalysis of older data revealed that source was already active on October 21 1988 \citep{Sunyaev1989}. We adopt this date as the start of the outburst and used the observational data from TTM \citep[\textit{Mir-Kvant,}][]{aleksandrovich2002} and ART-P \citep[\textit{GRANAT},][]{chelovekov2006}, which cover the first $\sim$7 years of the outburst. As of January 2006, the source was monitored with the Rossi X-Ray timing Explorer (\textit{RXTE}). We obtained monitoring data from its All Sky Monitor (ASM)\footnote{\scriptsize{\url{http://xte.mit.edu/ASM_lc.html}}} and pointed observations with the Proportional Counter Array (PCA)\footnote{\scriptsize{\url{http://asd.gsfc.nasa.gov/Craig.Markwardt/galscan/html/KS_1731-260.html}}}. The source was no longer detected with the ASM in January 2001 \citep{wijnands2001}, which was confirmed with pointed PCA observations and Galactic bulge scan observations \citep{markwardt2000, markwardt2000p}. The last PCA detection of the source was obtained on 2001 January 21 \citep{cackett2006}, and the first non-detection was on 2001 February 7 \citep{wijnands2001}. We adopt the latter date as end of the outburst, resulting in a total outburst duration of 12.3 years. 

We converted the ART-P and TTM mCrab-fluxes into daily averaged $2-10$ keV fluxes ($F(t)$), using a conversion factor 1 mCrab$=2.2\times10^{-11}$ erg cm$^{-2}$ s$^{-1}$. To convert the ASM ($1.5-12$ keV) and PCA ($2-60$ keV) count rates into $2-10$ keV fluxes we used WebPIMMS\footnote{\scriptsize{\url{http://heasarc.gsfc.nasa.gov/cgi-bin/Tools/w3pimms/w3pimms.pl}}}, assuming a spectral power law with photon index 2 and a Galactic absorption column density $n_\text{H}=1.3\times 10^{22}\text{ cm}^{-2}$ \citep{cackett2006}. Figure \ref{fig:lightcurve} shows the combined light curve. 

Next, we calculated a time-dependent mass accretion rate, $\dot{M}(t)$, using 
\begin{equation}
\dot{M}(t)=\frac{1.8F(t)4\pi d^2}{\eta c^2}
\label{mdot}
\end{equation}

\noindent where $c$ is the speed of light. We assumed a distance $d=7$ kpc \citep{muno2000}, a bolometric correction factor of 1.8, and took the fraction of the accreted mass that is converted into X-ray luminosity to be $\eta=0.2$. We note that there are large uncertainties in the published bolometric correction factors \citep[e.g.][]{zand2007,galloway2008}. Our conclusions are not influenced by the exact value that is used. 

We compute the accretion rate with a time accuracy of one day. If data from different telescopes overlap, we use PCA data if available; otherwise ASM data has priority over TTM data. For periods of missing data we estimate the accretion rate using linear extrapolation between the rates at the beginning and end of the data gap. This extrapolation is performed using one hour time resolution. The main reason for this accuracy is to obtain a high resolution estimate near the end of the outburst, as there are large gaps between the last two detections (61 days) and between the last detection and first non-detection (17 days). We obtain an average outburst accretion rate $\langle\dot{M}\rangle=0.095\dot{M}_\text{Edd}$ ($\dot{M}_\text{Edd}=1.58\times 10^{18}$ g s$^{-1}$).

\section{Results}
When KS 1731-260 went into quiescence, observational campaigns with \textit{XMM-Newton} and \textit{Chandra} tracked the neutron star crust cooling \citep{wijnands2001, wijnands2002, cackett2006, cackett2010}. Using our model, we obtain a crust cooling curve and compare this with the observed temperatures. The code leaves several free parameters which we adjust to obtain a best-fit of the observational data. 
\subsection{Time-dependent versus constant accretion rate}
The left panel in Figure \ref{fig:step_real} shows the calculated cooling curves from two different models. The red curve (model 1) shows our best-fit model when using a variable accretion rate as calculated from the observational data. Table \ref{tab:param} provides the parameters that we adopted in this model. The red curve in the inset in the figure shows the effective temperature as function of time during and after the outburst. The variations in temperature during the outburst are caused by variations in the accretion rate, as the outmost layers of the star (up to $\sim$$3\times10^{9}\text{ g cm}^{-3}$) have a nearly instantaneous response time, which can be observed from video 1 (see online material). 
\begin{table}
\centering
\caption{Best fit parameters for model 1.}
\label{tab:param}
\begin{tabular}{cccccccc}
\hline\hline
M & R & T$_0$    & y$_\text{light}$ & Q$_\text{sh}$ & $\rho_\text{sh,min}$ & {Q$_\text{imp}$} \\
(M$_\odot$) & (km) & (K)             &    (g cm$^{-2}$)            & (MeV)        & (g cm$^{-3}$)       &             \\ \hline
1.5         & 11.0 & 6.6$\times10^7$ & $10^6$  & 1.4          & 4$\times10^8$       & 0.6\\ \hline         
\end{tabular}
\end{table}

The black cooling curve in Figure \ref{fig:step_real} (model 2) shows the results of a model for which we have used the same input parameters as for model 1; the only difference being the fact that instead of a time-dependent accretion rate we have used a constant accretion rate (i.e. we assumed a step function as outburst profile). The accretion rate of model 2 is equal to the time-averaged accretion rate of model 1, $0.095\dot{M}_\text{Edd}$, such that the total amount of accreted material is the same for the two models. Figure \ref{fig:step_real} illustrates the strong effect of taking into account the observed variations in the light curve, especially in the first few hundred days of quiescence. The calculated cooling curve from model 2 does not provide a good fit to the observational data, except for the last few points. The parameters that are used in model 2 have to be modified to provide a good fit to the data points. Because the offset from the observational data is largest in the earliest phase of the cooling curve, this could be most easily achieved by decreasing the amount of shallow heating, $Q_\text{sh}$, to 0.6 MeV nuc$^{-1}$ (i.e. about halve the original value; see Table \ref{tab:param}). The amount of shallow heat regulates the temperature profile in the outer layers of the crust. Since the heat stored in the outermost layers starts diffusing outwards first after the end of the outburst, the temperature profile of the outer regions of the crust determine the initial part of the cooling curve \citep{brown2009}.

The right panel in Figure \ref{fig:step_real} shows the temperature profile at the end of the outburst for model 1 (red) and model 2 (black). Model 2 reaches a steady state $\sim$8 years after the start of the outburst (see video 1). However, if a time-dependent accretion rate is taken into account, KS 1731-260 never reaches a steady state profile due to the large accretion rate variations. Even the deepest crust layers ($10^{12-14}\text{ g cm}^{-3}$) are sensitive to variabilities on scales of years.
\subsection{Long time-scale outburst variability}
The amount of data that was obtained during the first $\sim$7 years of the outburst is limited, creating a high uncertainty on the estimated outburst accretion rates. Additionally, the exact start time of the outburst is unknown. To investigate the influence of long time-scale ($\gtrsim$1 yr) variations in the accretion rate in the first 7 years of the outburst on the calculated cooling curves, we created four models, each assuming a different accretion rate in the first part of the outburst (top panel Figure \ref{fig:smooth}). For all models, we assumed the same time-dependent accretion rate during the final $\sim$5 years of the outburst as in model 1. For model 3, we assumed no accretion during the first 7 years (which one might have assumed if no data was available at all during this period). For models 4, 5, and 6, we assumed a constant accretion rate at a level of 50\%, 100\%, and 200\% respectively, of the average accretion rate during this part of the outburst as determined from the data, which is $\dot{M}_\text{ini}=0.087\dot{M}_\text{Edd}$.

The four calculated cooling curves (Figure \ref{fig:smooth}; top panel) show small but significant differences that decrease until the crust has thermally relaxed ($\sim$$2\times10^3$ days after the end of the outburst). We have again assumed the same model parameters for the neutron star as in model 1 (see Table \ref{tab:param}). Since the difference in cooling curves is small for the considered accretion rates, this offset can be relatively easy compensated if one slightly adjusts one or two of the free parameters, with no explicit preference for which parameter(s).  
\subsection{Short time-scale outburst variability}

\begin{figure}	
     \includegraphics[width=\columnwidth]{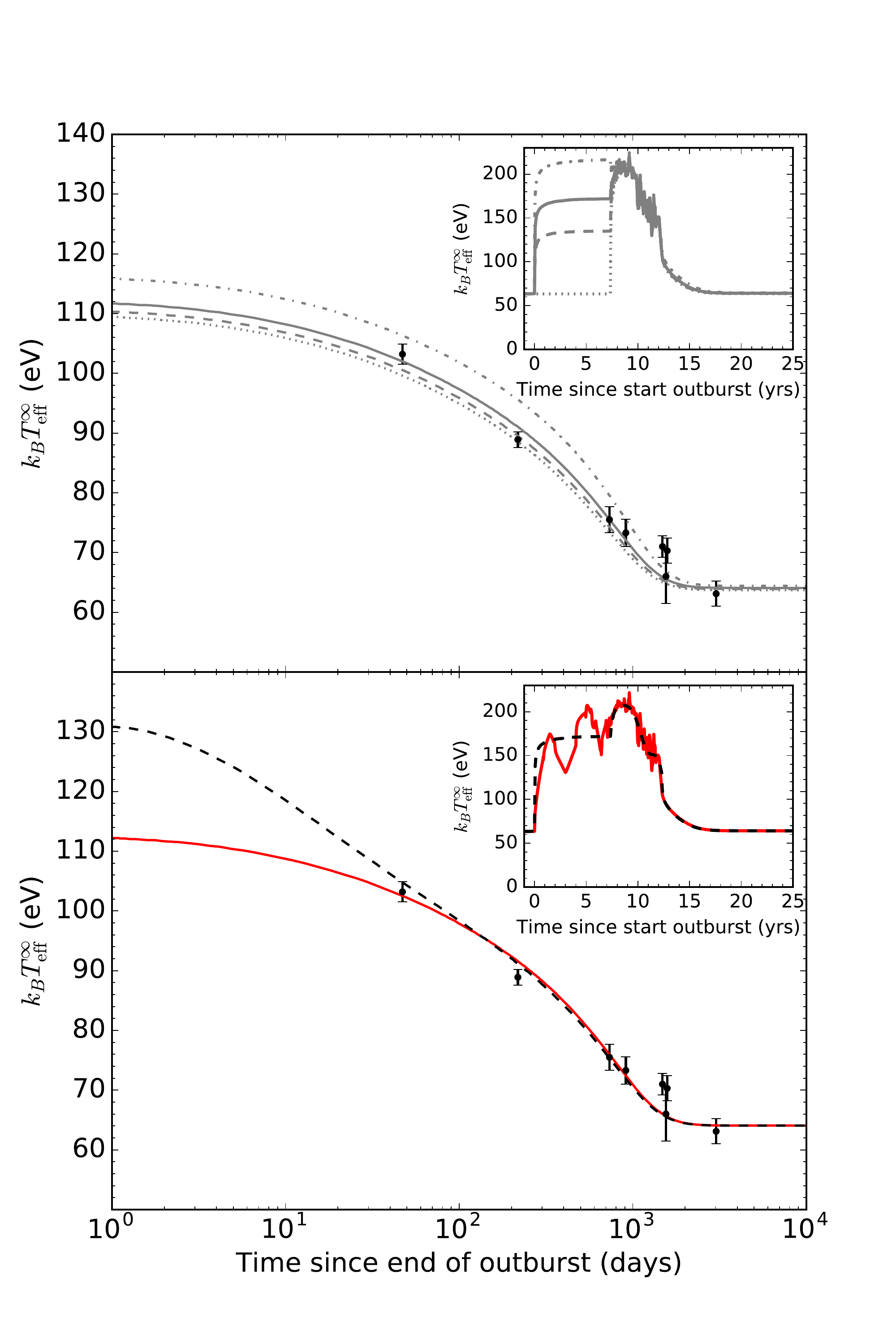}
     \vspace{-0.2cm}
     \caption{Same as Figure \ref{fig:step_real} left. Top: model 3 (dotted, initial $\dot{M}=0$), model 4 (dashed, initial $\dot{M}$=0.5$\langle\dot{M}_\text{ini}\rangle$, with $\dot{M}_\text{ini}$=0.087$\dot{M}_\text{Edd}$), model 5 (solid, initial $\dot{M}$=$\langle\dot{M}_\text{ini}\rangle$), and model 6 (dash-dotted, initial $\dot{M}$=2.0$\langle\dot{M}_\text{ini}\rangle$). Bottom: models 1 (red) and 7 (black dashed: smoothed $\dot{M}(t)$ last 5 yrs).}
     \label{fig:smooth}\vspace{-0.5cm}
\end{figure}

The calculated cooling curve from model 5 does not show significant differences from model 1, indicating that short time-scale (days-weeks) variability during the first part of the outburst does not affect the temperature profile in the neutron star crust at the end of the outburst. To determine the effect of short time-scale variability during the final part of the outburst on the cooling curve, we created model 7. In this model we smoothed out all variations on short time-scales during the last 5 outburst years, and for the initial part we use again a constant accretion rate equal to the average accretion rate obtained from the corresponding part of the light curve. 

The calculated cooling curve from model 7 provides a good fit of the observational data if we use the same model parameters as in model 1 (bottom panel Figure \ref{fig:smooth}). However, the effective temperatures during the first $\sim$40 days of quiescence are significantly higher than for model 1, because the smoothed fluxes provide a slight overestimation of the accretion rate during the decay phase of the outburst; in model 1 the accretion rate decreases from $0.026\dot{M}_\text{edd}$ to $0.006\dot{M}_\text{edd}$ in the last 78 days of the outburst, compared to a decrease from $0.038\dot{M}_\text{edd}$ to $0.026\dot{M}_\text{edd}$ in model 7. 

We tested for which period of time the short time-scale variations in accretion rate influence the cooling curve. For this specific source, we find that as long as the accretion rate during the last year of the outburst is determined from the obtained fluxes (while using the constant average accretion rate during the first $\sim$11 yrs), the calculated cooling curve does not differ significantly from the curve obtained from model 1, while for shorter periods it does. 

Additionally, we tested the influence of the light curve uncertainties in the last 78 days of the outburst (a 61-day gap between the last two detections and a subsequent 17-day gap until the first non-detection) using three scenarios: during the data gaps the accretion rate stayed constant compared to 1) the previous observation, 2) the next observation, and 3) the outburst ends directly after the last detection. We find that although all scenarios provide a good fit of the data if we use the same parameters as for model 1, the light curve uncertainties provide temperatures up to 4 eV lower and 6 eV higher than for model 1 at the start of quiescence. This is comparable to the typical 1$\sigma$ uncertainties from the observed surface temperatures.
\section{Discussion}
We set out to investigate to what accuracy the accretion outburst of KS 1731-260 has to be modelled to constrain the neutron star crustal parameters from the cooling curve predicted by crust cooling models. Comparison of our calculated cooling curves based on different input outburst profiles shows that it is important to model the accretion rate and its variability during outburst as accurately as possible. We compared the results of a model that uses a step function as outburst profile (constant accretion rate) with a model that uses a time-dependent accretion rate and find that the two can give strongly different cooling curves for the same crustal parameters up to $\sim$1000 days into quiescence. Since the difference is largest in the earliest phase of quiescence, the two models would likely lead to different constraints for the shallow heat source. All models of KS 1731-260 presented in this work require shallow heating to explain the large and rapid observed temperature decrease in quiescence. 

Comparing the temperature profiles at the end of the outburst for the two models shows that if a constant accretion rate is assumed, the source will reach a steady state within $\sim$8 years after the start of the outburst (see video 1), consistent with \citet{brown2009}. However, we find that a steady state is not reached when accretion variability is taken into account. This suggests that one cannot assume a steady state temperature profile as initial profile to calculate the cooling curve \citep{brown2009}.

We tested the effects of long time-scale (years) variabilities on the calculated cooling curves, because the accretion rate during the initial part of this outburst is highly uncertain due to the limited available observational data (see Figure \ref{fig:lightcurve}). We found that variations in the average accretion rate during the first seven years of the outburst have a small, but non negligible effect on the calculated cooling curve, indicating the importance of obtaining observations in all parts of the outburst. Because the differences in the cooling curves are constant throughout the cooling phase, uncertainties in the start time of the outburst and the average accretion rate during the first few years affect all crustal parameters. Consequently, if parts of the outbursts are missed it is essential that one models the time evolution of the temperature in the crust using different assumptions on the accretion rate during the missed parts. Albeit these effects are source-specific, uncertainties in the outburst profile of any source are expected to affect its cooling curve. 

Short time-scale (days-weeks) variability in the first few years of the outburst has no effect on the calculated cooling curves. On the other hand, short time-scale variations in the accretion rate during the final phase of the outburst, especially the last year, do strongly affect the outcome of the cooling curve for KS 1731-260. The outer layers of the neutron star have a very short thermal response time. Consequently, these layers already start to cool during the decay phase of the outburst, as can be clearly seen from video 1. As a result, the cooling curve starts at a significantly lower temperature when this effect is taken into account \citep[see also][]{deibel2015}. We find that even a small difference in modelling the decay phase of the outburst, and the uncertainties in the observational data of the last 78 days have a significant effect on the first $\sim$40 days of the calculated cooling curve. Since the amount of shallow heating sets the temperature in the outermost layers of the crust, this parameter is very sensitive to variations in the decay phase of the outburst. If the decay of the outburst is not taken into account properly, the accretion rate in this phase is overestimated and the calculated cooling temperatures are initially too high. Consequently, the amount of shallow heating is likely to be underestimated.

The latter result has important implications for constraining the origin of the shallow heat source. Constraining the amount and depth of shallow heat is key to solving this gap in our current understanding of neutron star crusts. Our results show that to constrain the strength of the shallow heat source from crust cooling models it is stringent to observe and model the decay phase of the outburst in detail. Additionally, observations in the earliest phase of quiescence (the first month) are required. 

Although our research only focusses on KS 1731-260, our findings apply to crust cooling models in general. For short-duration transients (outbursts of weeks-months), it is even more important to model variations in the outburst light curve, as these sources are further off from reaching a thermal steady state during outburst. \\

\noindent LO and RW acknowledge support from an NWO Top Grant, Module 1, awarded to RW.  DP is partially supported by the Mexican Conacyt (CB-2014-01, \#240512). ND acknowledges support from an NWO Vidi grant and an EU Marie Curie Intra-European fellowship. 



\vspace{-0.5cm}
\bibliographystyle{mnras}
\bibliography{Cooling_KS1731} 
\vspace{-0.2cm}








\bsp	
\label{lastpage}
\end{document}